\documentclass[showpacs,preprintnumbers]{revtex4}
\usepackage{amssymb}
\usepackage{graphicx}
\usepackage{color}

\def\bc{\begin{center}}
\def\ec{\end{center}}

\def\beq{\begin{equation}}
\def\eeq{\end{equation}}

\begin{document}

\title{High-temperature superfluidity of the two-component Bose gas in a TMDC bilayer}
\author{Oleg L. Berman$^{1,2}$ and Roman Ya. Kezerashvili$^{1,2}$}
\affiliation{\mbox{$^{1}$Physics Department, New York City College
of Technology, The City University of New York,} \\
Brooklyn, NY 11201, USA \\
\mbox{$^{2}$The Graduate School and University Center, The
City University of New York,} \\
New York, NY 10016, USA}
\date{\today}

\begin{abstract}

The high-temperature superfluidity of two-dimensional dipolar
excitons in two parallel TMDC layers is predicted.  We study
Bose-Einstein condensation in  the two-component system of dipolar A
and B excitons. The effective mass, energy spectrum of the
collective excitations, the sound velocity and critical temperature
are obtained for different TMDC materials. It is shown that in the
Bogolubov approximation the sound velocity in  the two-component
dilute exciton Bose gas is always larger than in any one-component.
The difference between the sound velocities for two-component and
one-component dilute gases is caused by the fact that the sound
velocity for two-component system depends on the reduced mass of A
and B excitons, which is always smaller than the individual mass of
A or B exciton. Due to this fact, the critical temperature $T_{c}$
for superfluidity for the two-component exciton system in TMDC
bilayer is about one order of magnitude higher than $T_{c}$ in any
one-component exciton system. We propose to observe the
superfluidity of two-dimensional dipolar excitons in two parallel
TMDC layers, which causes two opposite superconducting currents in
each TMDC layer.

\end{abstract}

\maketitle

\section{Introduction}

\label{intro}

The phenomenon known as Bose--Einstein condensation (BEC) occurs when a
substantial fraction of the bosons at low temperatures spontaneously occupy
the single lowest energy quantum state~\cite{Bose,Einstein}. The BEC can
cause the superfluidty in the system of bosons similarly to the superfluid
helium~\cite{Griffin,Pitaevskii}. A BEC of weakly interacting particles was
achieved experimentally in a gas of rubidium~\cite{Anderson,Ensher} and
sodium~\cite{Ketterle_Druten,Ketterle_Miesner} atoms. Cornell, Ketterle and
Wieman shared the 2001 Nobel Prize in Physics ``for the achievement of BEC
in dilute gases of alkali atoms''. The enormous technical challenges had to
be overcome in achieving the nanokelvin temperatures needed to create this
atomic BEC. The experimental and theoretical achievements in the studies of
the BEC of dilute supercold alkali gases are reviewed in Ref.~%
\onlinecite{Daflovo}.

Since the de Broglie wavelength for the two-dimensional (2D) system
is inversely proportional to the square root of the mass of a
particle, BEC can occur at much higher temperatures in a
high-density gas of small mass bosons, than for regular relatively
heavy alkali atoms. The very light bounded boson quasiparticles can
be produced using the absorption of a photon by a semiconductor
causing the creation of an electron in a conduction band and a
positively charge ``hole'' in a valence band. This electron-hole
pair can form a bound state
 known as an ``exciton''. The mass of an exciton is much smaller
than the mass of a regular atom. Therefore, such excitons are
expected to experience BEC and form superfluid at experimentally
observed exciton densities at temperatures much higher than for
alkali atoms~\cite{Moskalenko_Snoke}.

The prediction of superfluidity and BEC of dipolar (indirect) excitons
formed by spatially separated electrons and holes in semiconductor coupled
quantum wells (CQWs) attracted interest to this system~\cite%
{Lozovik,Littlewood,Vignale,Ulloa,Snoke,Butov,Eisenstein,BLSC,BKKL}.
In the CQWs negative electrons are trapped in a two-dimensional
plane, while an equal number of positive holes is located in a
parallel plane at a distance $D$ away. In this system the
electron-hole recombination due to the tunneling of electrons and
holes between different quantum wells is suppressed by the
dielectric barrier that separates the quantum wells. So the excitons
can have very long lifetime~\cite{Moskalenko_Snoke}, and, therefore,
they can be treated as metastable particles described by
quasiequilibrium statistics. At large enough separation distance $D$
the excitons experience the dipole-dipole repulsive interaction.

In the last decade many experimental and theoretical studies were devoted to
graphene, which is a 2D atomic plane of carbon atoms, known for unusual
properties in its band structure~\cite{Castro_Neto,Das_Sarma}. The
condensation of electron-hole pairs formed by spatially separated electrons
and holes in the two parallel graphene layers has been studied in Refs.~%
\onlinecite{Sokolik,Zhang,Min,Bistritzer,Efetov}. The excitons in gapped
graphene can be created by laser pumping. The superfluidity of
quasi-two-dimensional dipolar excitons in two parallel graphene layers in
the presence of band gaps was predicted recently in Ref.~\onlinecite{BKZ1}.

 Today an intriguing counterpart to gapless graphene is
a class of monolayer direct bandgap materials, namely transition
metal dichalcogenides (TMDCs). Monolayers of TMDC  such as
$\mathrm{Mo S_{2}}$, $\mathrm{Mo Se_{2}}$, $\mathrm{Mo Te_{2}}$, $\mathrm{W S_{2}}$, $%
\mathrm{W Se_{2}}$, and $\mathrm{W Te_{2}}$ are 2D semiconductors, (below
for TMDC monolayer we use the chemical formula $\mathrm{M X_{2}}$, where $%
\mathrm{M}$ denotes a transition metal $\mathrm{M}=\mathrm{Mo}$ or $\mathrm{W%
}$, and $\mathrm{X}$ denotes a chalcogenide, $\mathrm{X}=\mathrm{S}$, $%
\mathrm{Se}$ or $\mathrm{Te}$) which have the variety of
applications in electronics and opto-electronics~\cite{Kormanyos}.
The strong interest to the TMDC monolayers is caused by the
following facts: these materials have the direct gap in a
single-particle spectrum exhibiting the semiconducting band
structure~\cite{Mak2010,Mak2012,Novoselov,Zhao}, existence of
excitonic valley physics~\cite{Yao,Cao}, demonstration of strong
light-matter interactions that are electrically
tunable~\cite{Ross,Mak2013}.
The electronic band structure of TMDC monolayers was  calculated~\cite%
{Bromley3} by applying the semiempirical tight binding
method~\cite{Bromley1}
and the nonrelativistic augmented-plane-wave 
method~\cite{Mattheis}. The band structures and corresponding effective-mass
parameters have been calculated for bulk, monolayer, and bilayer TMDCs in
the $GW$ approximation, by solving the Bethe-Salpeter equation (BSE)~\cite%
{Lambrecht,Ashwin,Molina,Huser} and using the analytical approach~\cite%
{Malic}. The properties of direct excitons in mono- and few-layer TMDCs on a
$\mathrm{SiO_{2}}$ substrate were experimentally and theoretically
investigated, identifying and characterizing not only the ground-state
exciton but the full sequence of excited (Rydberg) exciton states~\cite%
{Chernikov}. The exciton binding energy for monolayer, few-layer and bulk
TMDCs and optical gaps were evaluated using the tight-binding approximation~%
\cite{MacDonald}, by solving the BSE~\cite{Komsa,Yakobson}, applying
an effective mass model, density functional theory  and subsequent
random
phase approximation 
calculations~\cite{Reichman}, and by generalized time-dependent
density-matrix functional theory approach~\cite{Rahman}. Significant
spin-orbit splitting in the valence band leads to the formation of two
distinct types of excitons in TMDC layers, labeled A and B~\cite{Reichman}.
The excitons of type A are formed by spin-up electrons from conduction and
spin-down holes from valence bands. The excitons of type B are formed by
spin-down electrons from conduction and spin-up holes from valence bands.
According to Figure~4 in Ref.~\onlinecite{Kormanyos}, the spin-orbit
splitting in the valence band is much larger than in the conduction band.
 For both $\mathrm{Mo X_{2}} $ and $\mathrm{W X_{2}}$ in the valence band the energy
for spin-down electrons is larger than for spin-up electrons. The
spin-orbit spitting causes the experimentally observed energy
difference between the A and B excitons~\cite{Kormanyos}. Two-photon
spectroscopy of excitons in monolayer TMDCs was studied using a
BSE~\cite{Hybertsen}.

 Recently it was proposed a design of the
heterostructure of two TMDC monolayers, separated by a hexagonal
boron nitride (hBN) insulating barrier for observation of a high
temperatures superfluidity~\cite{Fogler}.   The emission of neutral
and charged excitons was controlled by the gate voltage,
temperature, the helicity and the power of optical excitation. The
formation
of indirect excitons in a heterostructure formed in monolayers of $\mathrm{%
Mo S_{2}}$ and $\mathrm{Mo Se_{2}}$ on a
$\mathrm{Si}-\mathrm{SiO_{2}}$ substrate was
observed~\cite{Ceballos}. The dynamics of direct and indirect
excitons in $\mathrm{W Se_{2}}$ bilayers was studied experimentally
applying time-resolved photoluminescence
spectroscopy~\cite{Wang_apl}. We propose the theoretical description
for the superfluidity of two-component Bose gas of such dipolar
excitons in various TMDC bilayers.

 The important peculiarity of the system of dipolar
excitons in TMDC bilayer is caused by the fact that this system is a
two-component mixture of A and B excitons.  The two-component
mixtures of trapped cold atoms experiencing BEC and superfluidity
have been  the subject of various experimental and theoretical
studies~\cite{Jin,Ohberg,Altman,Tsubota}. The Hamiltonian of
two-component Bose systems includes the terms, corresponding to
three types of interactions: the interaction between the same bosons
for both species and the interaction between the different bosons
from the different species. These three interaction terms in the
Hamiltonian are described by three different interaction constants.
The Bogoliubov approximation was applied to describe the excitation
spectrum of two-component BEC of cold
atoms~\cite{Timmermans,Passos,Sun}. We apply the Bogoliubov
approximation, developed for two-component atomic BEC, to derive the
excitation spectrum of two-component BEC  of A and B dipolar
excitons in a TMDC bilayer.

In this Paper we consider the dilute gas of dipolar excitons formed
by an electron and a hole in two parallel spatially separated TMDC
monolayers. The spatial separation of electrons and holes in
different monolayers results in increasing of the exciton life time
compare to direct excitons in a single monolayer due to small
probability of the tunneling between monolayers, since the
monolayers are separated by the dielectric barrier. We consider the
formation of a BEC for A and B dipolar excitons that are in the
ground state. To find the single-particle spectrum for a single
dipolar exciton we solve analytically the two-body problem for a
spatially separated electron and a hole located in two parallel
 TMDC layers.  The last allows us to obtain the spectrum
of the collective excitations and the sound velocity for a dilute
two-component exciton Bose gas formed by A and B excitons within the
framework of the Bogoliubov approximation. The superfluid phase can
be formed at finite temperatures due to the dipole-dipole
interactions between dipolar excitons, which result in the sound
spectrum at small momenta for the collective excitations. The sound
spectrum satisfies to the Landau criterion of the superfluidity~\cite%
{Abrikosov,Lifshitz}. We calculated the spectrum of collective
excitations, the density of a superfluid component as a function of
temperature, and the mean field phase transition temperature, below
which superfluidity occurs in this system. We predict the existence
a high-temperature superfluidity of dipolar excitons in two TMDC
layers at the temperatures below the mean field phase transition
temperature. Our most fascinating finding is that in the Bogolubov
approximation the sound velocity in a two-component dilute Bose gas
of indirect excitons is always larger than in any one-component Bose
gas in CQWs and that leads to a remarkable high-temperature
superfluidity.

The paper is organized in the following way. In Sec.~\ref{2body}, we
solve the eigenvalue problem for an electron and a hole in two
different parallel TMDC layers, separated by a dielectric. The
effective masses and the single-particle energy spectra of the
dipolar excitons in two parallel TMDC layers  are obtained. In
Sec.~\ref{collect}, we study the condensation of  the two-component
gas of dipolar A and B excitons and calculate the spectrum of
collective excitations. In Sec.~\ref{sup} we obtain the density of
the superfluid component as well as the mean field phase transition
temperature. The specific properties of the superfluid of direct
excitons in a TMDC monolayer are discussed in Sec.~\ref{directex}.
The results of the
calculations and their discussion  are presented in Sec.~\ref%
{disc}. The conclusions follow in Sec.~\ref{conc}.

\section{Two-body problem for Dirac particles with a gap}

\label{2body}

\begin{figure}[h]
\includegraphics[width=12.0cm]{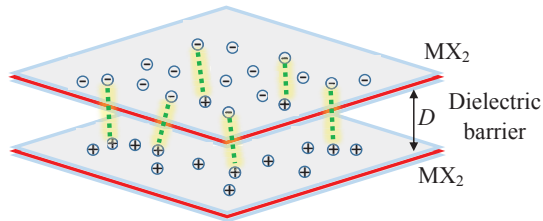} \vspace{-12cm}
\caption{Spatially separated electrons and holes in two monolayers of TMDC.}
\label{twolayers}
\end{figure}

\begin{figure}[h]
\includegraphics[width=12.0cm]{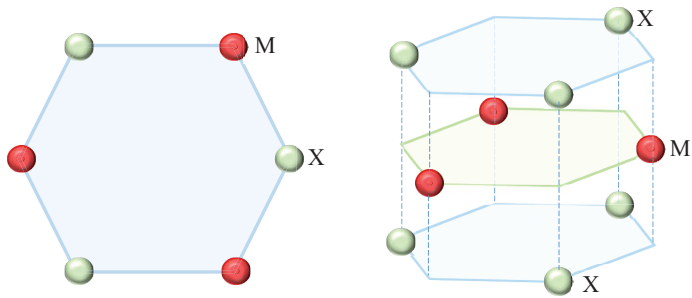} \vspace{-10cm}
\caption{The structure of a TMDC monolayer.}
\label{structureTMDC}
\end{figure}

The formation of excitons in two parallel graphene layers separated
by an insulating material due to gap opening in the electron and
hole spectra in the two graphene layers was considered in
Ref.~\onlinecite{BKZ1}. Here we apply the similar approach to study
excitons in coupled quantum wells designed from atomically thin
materials stacked on top of each other and separated by a dielectric
barrier. Let us consider indirect excitons composed by
 electrons and holes located in two different parallel TMDC
monolayers separated by an insulating barrier of a thickness $D$ as
shown  in Fig.~\ref{twolayers}.
Each monolayer TMDC has hexagonal lattice structure and consists of an
atomic layer of a transition metal $\mathrm{M}$ sandwiched between two
layers of a chalcogenide $\mathrm{X}$ in a trigonal prismatic structure as
shown in Fig.~\ref{structureTMDC}.

In TMDC materials the physics around the $K$ and $-K$ points has
attracted the most attention both experimentally and theoretically.
Today the gapped Dirac Hamiltonian model, that contains only the
terms linear in $p$ and the spin-splitting in the valence band, is
widely used~\cite{Yao}. The low-energy effective two-band single
electron Hamiltonian in the form of a spinor with a gapped spectrum
for TMDCs in the $k \cdot p$ approximation is given by~\cite{Yao}
\begin{eqnarray}  \label{Hamyao}
\hat{H}_{s} = at\left(\tau k_{x} \hat{\sigma}_{x} + k_{y}\hat{\sigma}%
_{y}\right)+ \frac{\Delta}{2}\hat{\sigma}_{z} - \lambda \tau \frac{\hat{%
\sigma}_{z}-1}{2}\hat{s}_{z} \ .
\end{eqnarray}%
In Eq.~(\ref{Hamyao}) $\hat{\sigma}$ denotes the Pauli matrices, $a$
is the lattice constant, $t$ is the effective hopping integral,
$\Delta$ is the energy gap, $\tau = \pm 1$ is the valley index,
$2\lambda$ is  the spin splitting at the valence band top caused by
the spin-orbit coupling (SOC), and $\hat{s}_{z}$ is the Pauli matrix
for spin that remains a good quantum number. The parameters of the
Hamiltonian $\hat{H}_{s}$ presented by Eq.~(\ref{Hamyao}) for
transition metal dichalcogenides $\mathrm{Mo S_{2}}$, $\mathrm{Mo
Se_{2}}$, $\mathrm{W S_{2}}$, and $\mathrm{W Se_{2}}$ are listed
in Refs.~\onlinecite{Kormanyos,Yao}, and in Ref.~\onlinecite{Kormanyos} the parameters for  $\mathrm{Mo Te_{2}}$ and $%
\mathrm{W Te_{2}}$ are presented.

We consider two  parallel TMDC layers with the interlayer separation
$D$. The dipolar excitons in this double-layer system are formed by
the electrons located in one TMDC layer, while the holes located in
another one. Let us mention that the electron moves in one TMDC
layer, and the hole moves in the other TMDC layer. So the coordinate
vectors of the electron and hole can be replaced by their 2D
projections on plane of one of
the TMDC layer. These new in-plane coordinates $\mathbf{r}_{1}$ and $\mathbf{%
r}_{2}$ for an electron and a hole, correspondingly, will be used
everywhere below. In each TMDC layer a quasiparticle is
characterized by the coordinates $\mathbf{r}_{j}$ in the
conduction ($c$) and valence ($v$) band with the corresponding direction of spin ($s_{j})$ up $%
\uparrow $ or down $\downarrow $, and index $j=1,2$ referring to the two
monolayers, one with electrons and the other with holes. The spinful basis
for description of two particles in different monolayers is given by $%
\left\{ \left\vert \Psi _{jc},s_{jc}\right\rangle ,\left\vert \Psi
_{jv},s_{jv}\right\rangle \right\} $, where $\left\vert \Psi
_{jc},s_{jc}\right\rangle =\left\vert \Psi _{jc}\right\rangle
\otimes \left\vert s_{jc}\right\rangle $ and $\left\vert \Psi
_{jv},s_{jv}\right\rangle =\left\vert \Psi _{jv}\right\rangle
\otimes
\left\vert s_{jv}\right\rangle $ with the coordinate wave functions $%
\left\vert \Psi _{jc}\right\rangle $ and $\left\vert \Psi
_{jv}\right\rangle $ and spin wave functions $\left\vert
s_{jc}\right\rangle $ and $\left\vert s_{jv}\right\rangle ,$ where
$s=\left\{ \uparrow ,\downarrow \right\} $   is denoting the spin
degree of freedom, in the conduction and valence bands for the first
and second monolayers, correspondingly.  Therefore, the two-particle
wave function that describes the bound electron and hole in
different monolayers,  reads $\Psi
_{s}(\mathbf{r}_{1},\mathbf{r}_{2})$.  This  wave function can also
be understood as a four-component spinor, where the spinor
components refer to the four possible values of the
conduction/valence band indices:

\begin{eqnarray}
\Psi _{\uparrow }(\mathbf{r}_{1}, \mathbf{r}_{2})=\left( {%
\begin{array}{c}
\phi _{c\uparrow c\uparrow }(\mathbf{r}_{1},\mathbf{r}_{2}) \\
\phi _{c\uparrow v\uparrow }(\mathbf{r}_{1},\mathbf{r}_{2}) \\
\phi _{v\uparrow c\uparrow }(\mathbf{r}_{1},\mathbf{r}_{2}) \\
\phi _{v\uparrow v\uparrow }(\mathbf{r}_{1},\mathbf{r}_{2})%
\end{array}%
}\right) \equiv \left( {%
\begin{array}{c}
\Psi _{c\uparrow } \\
\Psi _{v\uparrow }%
\end{array}%
}\right) ,\text{ where }\Psi _{c\uparrow }=\left( {%
\begin{array}{c}
\phi _{c\uparrow c\uparrow } \\
\phi _{c\uparrow v\uparrow }%
\end{array}%
}\right) ,\ \ \ \Psi _{v\uparrow }=\left( {%
\begin{array}{c}
\phi _{v\uparrow c\uparrow } \\
\phi _{v\uparrow v\uparrow }%
\end{array}%
}\right)  \ .
 \label{wave function1}
\end{eqnarray}%
The two components reflect one particle being in the conduction
(valence) band and the other particle being in the valence
(conduction) band, correspondingly. Let us mention that while
Eq.~(\ref{wave function1}) represents the spin-up particles, the
spin-down particles  are represented by the same expression
replacing $\uparrow $ by $\downarrow $.

Each TMDC layer has an energy gap. Following the procedure applied for
double-layer gapped graphene in Ref.~\onlinecite{BKZ}, the Hamiltonian $%
H_{\uparrow (\downarrow )}$ for spin-up (spin-down) particles can be written
as
\begin{eqnarray}
\mathcal{H}_{\uparrow (\downarrow )}=\left(
\begin{array}{cccc}
V(r) & d_{2} & d_{1} & 0 \\
d_{2}^{\dagger } & -\Delta ^{\prime }+V(r) & 0 & d_{1} \\
d_{1}^{\dagger } & 0 & \Delta ^{\prime }+V(r) & d_{2} \\
0 & d_{1}^{\dagger } & d_{2}^{\dagger } & V(r)%
\end{array}%
\right) \ ,
 \label{k20}
\end{eqnarray}%
where $V(r)$ is the potential energy of the attraction between an
electron and a hole, the parameter $\Delta ^{\prime }$ is defined as
$\Delta ^{\prime }=\Delta -\lambda $ for spin-up particles, and
$\Delta ^{\prime }=\Delta +\lambda $ for spin-down particles. In
Eq.~(\ref{k20}) $d_{1}=at(-i\partial _{x_{1}}-\partial _{y_{1}})$,
$d_{2}=at(-i\partial _{x_{2}}-\partial _{y_{2}})$ and the
corresponding Hermitian conjugates are $d_{1}^{\dagger
}=at(-i\partial _{x_{1}}+\partial _{y_{1}})$, $d_{2}^{\dagger
}=at(-i\partial _{x_{2}}+\partial _{y_{2}})$, where $\partial
_{x}=\partial
/\partial x$ and $\partial _{y}=\partial /\partial y,$ $x_{1}$, $y_{1}$ and $%
x_{2}$, $y_{2}$ are the coordinates of vectors $\mathbf{r}_{1}$ and $\mathbf{%
r}_{2}$, correspondingly.

The single-particle energy spectrum of an electron-hole pair can be found by
solving the eigenvalue problem for Hamiltonian~(\ref{k20}):
\begin{eqnarray}
\mathcal{H}_{\uparrow (\downarrow )}\Psi _{\uparrow (\downarrow
)}=\epsilon _{\uparrow (\downarrow )}\Psi _{\uparrow (\downarrow )}\
,
 \label{l}
\end{eqnarray}
where $\Psi _{\uparrow (\downarrow )}$ are four-component eigenfunctions as
given in Eq.~(\ref{wave function1}), and $\epsilon _{\uparrow (\downarrow )}$
is the single-particle energy spectrum for an electron-hole pair with the up
and down spin orientation, correspondingly. In this notation we assume that
a spin-up (-down) hole describes the absence of a spin-down (-up) valence
electron.

For Hamiltonian~(\ref{k20}) the center-of-mass motion cannot be
separated from the relative motion due the chiral nature of Dirac
electron in TMDC. The similar conclusion was made for the
two-particle problem in graphene in Ref.~\onlinecite{Sabio} and
gapped graphene in Ref.~\onlinecite{BKZ}. Since the electron-hole
Coulomb interaction depends only on the relative coordinate, we
introduce the new ``center-of-mass'' coordinates in the plane of a
TMDC layer:
\begin{eqnarray}  \label{exp1}
\mathbf{R}=\alpha \mathbf{r}_{1}+\beta \mathbf{r}_{2}\ ,  \nonumber \\
\mathbf{r}=\mathbf{r}_{1} - \mathbf{r}_{2}\ ,
\end{eqnarray}
were the coefficients $\alpha $ and $\beta $ are supposed to be
found below from the condition of the separation of the coordinates
of the center-of-mass and relative motion in the Hamiltonian in the
one-dimensional equation for the corresponding component of the wave
function.

We make the following Anz\"{a}tze to obtain the solution of Eq.~(\ref{l})
\begin{eqnarray}
\Psi_{j\uparrow(\downarrow)}(\mathbf{R},\mathbf{r})=\mathtt{e}^{i\mathbf{K}%
\cdot \mathbf{R}}\psi _{j\uparrow(\downarrow)}(\mathbf{r})\ ,
\end{eqnarray}%
and  follow the procedure described for the two-body problem in
double-layer gapped graphene in Ref.~\onlinecite{BKZ}. The solution of a two-particle problem is demonstrated in Appendix~\ref%
{app:A}. Finally Eq.~(\ref{25}) that describes the bound electron-hole
system can be written in the following form
\begin{eqnarray}  \label{26}
\left( -F_{1}(\epsilon _{\uparrow (\downarrow )})\nabla _{\mathbf{r}%
}^{2}+V(r)\right) \phi _{c\uparrow (\downarrow )v\uparrow (\downarrow
)}=F_{0}^{\prime }(\epsilon _{\uparrow (\downarrow )})\phi _{c\uparrow
(\downarrow )v\uparrow (\downarrow )}\ ,
\end{eqnarray}
where
\begin{eqnarray}  \label{F01}
F_{1}(\epsilon _{\uparrow (\downarrow
)})=\frac{2a^{2}t^{2}}{\epsilon _{\uparrow (\downarrow )}}\
,\hspace{5cm}F_{0}^{\prime }(\epsilon _{\uparrow
(\downarrow )})=\epsilon _{\uparrow (\downarrow )}+\Delta ^{\prime }-\frac{%
a^{2}t^{2}\mathcal{K}^{2}}{2\epsilon _{\uparrow (\downarrow )}}\ .
\end{eqnarray}

We consider spatially separated an electron and a hole in two parallel TMDC
layers at large distances $D \gg a_{B}$, where $a_{B}$ is a 2D Bohr radius
of a dipolar exciton. For TMDC materials the Bohr radius of the dipolar
exciton is found to be in the rangde from $1.5 \ \mathrm{{\mathring{A}}}$
for $\mathrm{Mo Te_{2}}$~\cite{Ashwin} up to $3.9 \ \mathrm{{\mathring{A}}}$
for $\mathrm{Mo S_{2}}$~\cite{Louie}.

It is obvious that the electron and hole are interacting via the
Coulomb potential. However,~in general, the electron-hole
interaction is effected by the screening effects~\cite{Reichman}.
However,  the screening effects are negligible at long range for
electron-hole distances larger than the screening length $\rho
_{0}$, and at long range the electron-hole interaction is described
by the Coulomb's potential~\cite{Reichman}. The screening length is
defined as $\rho _{0}=2\pi \chi _{2D}$, where $\chi _{2D} $ is the
2D polarizability of the planar material~\cite{Rubio}. Substituting
$\chi _{2D}$ from Ref.~\onlinecite{Reichman}, we conclude that
for TMDC $\rho _{0}$ is estimated as $38\ \mathrm{{\mathring{A}}}$ for $%
\mathrm{WS_{2}} $, $41\ \mathrm{{\mathring{A}}}$ for $\mathrm{MoS_{2}}$, $%
45\ \mathrm{{\mathring{A}}}$ for $\mathrm{WSe_{2}}$, $52\ \mathrm{{\mathring{%
A}}}$ for $\mathrm{MoSe_{2}}$. The binding energy for the dipolar exciton
was estimated for two $\mathrm{MoS_{2}}$ layers separated by $N$ hBN
insulating layers from $N=1$ up to $N=6$~\cite{Fogler}. These dipolar
excitons were observed experimentally for $N=2$~\cite{Calman}. The
interlayer separation $D$ is given by $D=Nc$, where $c=0.333\ \mathrm{nm}$~%
\cite{Fogler}. We assume that the indirect excitons in TMDC can
survive for a larger interlayer separation $D$ than in semiconductor
coupled quantum
wells, because the thickness of a TMDC layer is fixed (for example, for $%
\mathrm{MoS_{2}}$ this thickness is $0.312\ \mathrm{nm}$~\cite{Fogler}),
while the spatial fluctuations of the thickness of the semiconductor quantum
well effect the structure of the dipolar exciton.

Since for the TMDCs materials the characteristic values for the 2D
exciton Bohr radius are found to be much less than the
characteristic values of the screening length $\rho _{0}$, Coulomb's
potential describes the electron hole interaction for $D>\rho _{0}$.
Otherwise, for $D\lesssim \rho _{0}$,  the electron-hole interaction
is described by Keldysh's potential due to the screening
effects~\cite{Keldysh}. Though  the two-body electron-hole problem
with Keldysh's potential can be solved only numerically, it cannot
be solved analytically. We solve the two-body electron-hole problem
analytically for  large interlayer distances $D>\rho _{0}$. In this
case, when the screening effects for the interaction between an
electron and a hole at large distances are negligible, the potential
energy $V(r)$ corresponding to the attraction between an electron
and a hole is given by
\begin{eqnarray}  \label{V}
V(r)=-\frac{ke^{2}}{\epsilon _{d}\sqrt{r^{2}+D^{2}}}\ ,
\end{eqnarray}
where $k=9\times 10^{9}\ N\times m^{2}/C^{2}$, $\epsilon _{d}$ is
the dielectric constant of the dielectric, which separates two TMDC
layers. Assuming $r\ll D$, we approximate $V(r)$ by the first two
terms of the Taylor series, and substituting
\begin{eqnarray}  \label{Vap}
V(r)=-V_{0}+\gamma r^{2}\ ,
\end{eqnarray}
where
\begin{eqnarray}  \label{V0g}
V_{0}=\frac{ke^{2}}{\epsilon _{d}D}\ ,\hspace{5cm}\gamma =\frac{ke^{2}}{%
2\epsilon _{d}D^{3}}\ ,
\end{eqnarray}
into Eq.~(\ref{26}), one obtains the equation in the form of
Schr\"{o}dinger equation for the 2D isotropic harmonic oscillator:
\begin{eqnarray}  \label{harm1}
\left( -F_{1}(\epsilon _{\uparrow (\downarrow )})\nabla _{\mathbf{r}%
}^{2}+\gamma r^{2}\right) \phi _{c\uparrow (\downarrow )v\uparrow
(\downarrow )}=F_{0}(\epsilon _{\uparrow (\downarrow )})\phi _{c\uparrow
(\downarrow )v\uparrow (\downarrow )}\ ,
\end{eqnarray}
where
\begin{eqnarray}  \label{F0111}
F_{0}(\epsilon _{\uparrow (\downarrow )})=\epsilon _{\uparrow
(\downarrow )}+\Delta ^{\prime
}+V_{0}-\frac{a^{2}t^{2}\mathcal{K}^{2}}{2\epsilon _{\uparrow
(\downarrow )}}\ .
\end{eqnarray}

The solution of the Schr\"{o}dinger equation for the harmonic
oscillator, is well known  and is given by
\begin{eqnarray}  \label{sol2D1}
\frac{\mathcal{F}_{0}(\epsilon_{\uparrow(\downarrow)} )}{\mathcal{F}%
_{1}(\epsilon_{\uparrow(\downarrow)} )}=2N\sqrt{\frac{\gamma }{\mathcal{F}%
_{1}(\epsilon_{\uparrow(\downarrow)} )}}\ ,
\end{eqnarray}
where $N=2\tilde{N}+|L|+1$, and $\tilde{N}=\mathrm{min}(\widetilde{n},%
\widetilde{n}^{\prime })$, $L=\widetilde{n}-\widetilde{n}^{\prime }$, $%
\widetilde{n},$ $\widetilde{n}^{\prime }=0,1,2,3,\ldots $ are the
quantum numbers of the 2D harmonic oscillator. The corresponding 2D
wave function at $\mathcal{K} = 0$ in terms of associated Laguerre
polynomials can be written as
\begin{eqnarray}
\phi _{c\uparrow(\downarrow)v\uparrow(\downarrow)\tilde{N}\text{ }L, \mathcal{K} = 0}(r)=%
\frac{\tilde{N}!}{a_{B}^{|L|+1}\sqrt{\widetilde{n}!\widetilde{n}^{\prime }!}}%
\mathrm{sgn}(L)^{L}r^{|L|-1/2}e^{-r^{2}/(4a_{B}^{2})}\times L_{\tilde{N}%
}^{|L|}(r^{2}/(2a_{B}^{2}))\frac{e^{-iL\phi }}{(2\pi )^{1/2}}\ ,
\label{rk14}
\end{eqnarray}%
where $\phi $ is the polar angle, $L_{k}^{p}(x)$ are the associated
Laguerre polynomials. and the Bohr radius of the dipolar exciton
$a_{B}$ is given by
\begin{eqnarray}  \label{Bohr1}
a_{B}= \left( \sqrt{F_{1}(\epsilon )}/\left( 2\sqrt{\gamma }\right)
\right)^{1/2} = \left(\frac{at}{ \sqrt{2\gamma \left|\epsilon\right|}}%
\right)^{1/2} \ .
\end{eqnarray}

Substituting Eqs.~(\ref{F01}) and~(\ref{F0111}) into Eq.~(\ref{sol2D1}), we
obtain
\begin{eqnarray}  \label{geneq}
2\epsilon_{\uparrow(\downarrow)}^{2} + 2\left(\Delta^{\prime} + V_{0}\right)
\epsilon_{\uparrow(\downarrow)} - \frac{8atN\sqrt{\gamma
\epsilon_{\uparrow(\downarrow)}}}{\sqrt{2}} - a^{2} t^{2} \mathcal{K}^{2} =
0 \ .
\end{eqnarray}

The solution of Eq.~(\ref{geneq}) for the single exciton spectrum is shown
in Appendix~\ref{app:B}. From Eq.~(\ref{exnm}), for the single exciton
spectrum one obtains
\begin{eqnarray}  \label{ex}
\epsilon _{A(B)}=x_{0}^{2}+\frac{\hbar
^{2}\mathcal{K}^{2}}{2M_{A(B)}}\ ,
\end{eqnarray}
where $M_{A(B)}$ is the dipolar exciton effective mass given by
\begin{eqnarray}  \label{mex}
M_{A(B)}=\frac{C_{A(B)}\hbar ^{2}}{2a^{2}t^{2}x_{0}}\ ,
\end{eqnarray}
where in Eqs.~(\ref{ex}) and~(\ref{mex}) $x_{0}$ has the different
value for A and B excitons.

The dipolar exciton binding energy is given by
\begin{eqnarray}  \label{Eb}
E_{b\ A(B)}=-\left( x_{0}^{2}-\Delta ^{\prime }\right) \ .
\end{eqnarray}
In Eq.~(\ref{Eb}), we assume $\Delta ^{\prime }=\Delta -\lambda $
for A excitons, and $\Delta ^{\prime }=\Delta +\lambda $ for B
excitons.


\section{The collective excitations for spatially separated electrons and
holes}

\label{collect}

Let us consider the dilute limit for the electrons and holes gases
in parallel TMDC layers spatially separated by the dielectric, when
$n_{A}a_{B\ A}^{2}\ll 1$ and $n_{B}a_{B\ B}^{2}\ll 1$, where
$n_{A(B)}$ and $a_{B\ A(B)}$ are the concentration and effective
exciton Bohr radius for A(B) dipolar
excitons, correspondingly. In the experiments, the exciton density in a $%
\mathrm{WSe_{2}}$ monolayer was obtained up to $n=5\times 10^{11}\ \mathrm{%
cm^{-2}}$~\cite{You_bi}. In the dilute limit, the dipolar A and B excitons
are formed by the electron-hole pairs with the electrons and holes spatially
separated in two different TMDC layers. The Hamiltonian $\hat{H}$ of the 2D
A and B interacting dipolar excitons is given by
\begin{eqnarray}  \label{Ham}
\hat{H}=\hat{H}_{A}+\hat{H}_{B}+\hat{H}_{I}\ ,
\end{eqnarray}
where $\hat{H}_{A(B)}$ are the Hamiltonians of A(B) excitons given
by
\begin{eqnarray}  \label{Ham1}
\hat{H}_{A(B)}=\sum_{\mathbf{k}}E_{A(B)}(k)a_{\mathbf{k}A(B)}^{\dagger }a_{%
\mathbf{k}A(B)}+\frac{g_{AA(BB)}}{2S}\sum_{\mathbf{k}\mathbf{l}\mathbf{m}}a_{%
\mathbf{k}A(B)}^{\dagger }a_{\mathbf{l}A(B)}^{\dagger }a_{A(B)\mathbf{m}%
}a_{A(B)\mathbf{k}+\mathbf{l}-\mathbf{m}}\ ,
\end{eqnarray}
and $\hat{H}_{I}$ is the Hamiltonian of the interaction between A
and B excitons given by
\begin{eqnarray}  \label{Ham2}
\hat{H}_{I}=\frac{g_{AB}}{S}\sum_{\mathbf{k}\mathbf{l}\mathbf{m}}a_{\mathbf{k%
}A}^{\dagger }a_{\mathbf{l}B}^{\dagger }a_{B\mathbf{m}}a_{A\mathbf{k}+%
\mathbf{l}-\mathbf{m}}\ ,
\end{eqnarray}
where $a_{\mathbf{k}A(B)}^{\dagger }$ and $a_{\mathbf{k}A(B)}$ are
Bose creation and annihilation operators for A(B) dipolar excitons
with the wave
vector $\mathbf{k}$, correspondingly, $S$ is the area of the system,  $%
E_{A(B)}(k)\equiv \epsilon _{A(B)} = \varepsilon
_{(0)A(B)}(k)+\mathcal{A}_{A(B)}$ is the energy
spectrum of non-interacting A(B) dipolar excitons, correspondingly,  $%
\varepsilon _{(0)A(B)}(k)=\hbar ^{2}k^{2}/(2M_{A(B)})$, $M_{A(B)}$
is an effective mass of non-interacting dipolar excitons,
$\mathcal{A}_{A(B)}$ is the constant, which depends  on   A(B)
dipolar exciton binding energy and the gap, formed by a spin-orbit
coupling for the A(B) dipolar exciton, $g_{AA(BB)} $ and $g_{AB}$
are the interaction constants for the interaction between two  A
dipolar excitons, two  B dipolar excitons with the same conduction
band electron spin orientation and for the interaction between A and
B dipolar excitons with the opposite conduction band electron spin
orientation.

We consider the dilute system, when the average distance between the
excitons is much larger than the interlayer separation $D$, which
corresponds to the densities $n\ll 1/(\pi D^{2})$. Since we assume that $%
D>\rho _{0}$, the screening effects are negligible, and the interaction
between the particles is described by the Coulomb's potential. For example,
for $D=50\ \mathrm{{\mathring{A}}}$, the exciton densities should be $n\ll
1.3\times 10^{12}\ \mathrm{cm^{-2}}$.

In the dilute system at the large interlayer separation $D$, two
dipolar excitons at the distance $R$ repel due to the dipole-dipole
interaction potential $U(R)=ke^{2}D^{2}/(\epsilon _{d}R^{3})$.
Following the procedure presented in Ref.~\onlinecite{BKKL}, the
interaction parameters for the exciton-exciton interaction in very
dilute systems could be obtained assuming the exciton-exciton
dipole-dipole repulsion exists only at the distances between
excitons greater than distance from the exciton to the classical
turning point. The distance between two excitons cannot be less than
this distance, which is determined by the conditions reflecting the
fact that the energy of two excitons cannot exceed doubled chemical
potential of the system $\mu $:
\begin{eqnarray}  \label{cond}
2\mathcal{A}_{A}+U(R_{0AA})=2\mu \ ,\hspace{1.5cm}2\mathcal{A}%
_{B}+U(R_{0BB})=2\mu \ ,\hspace{1.5cm}\mathcal{A}_{A}+\mathcal{A}%
_{B}+U(R_{0AB})=2\mu \ ,
\end{eqnarray}
where $R_{0AA}$, $R_{0BB}$, and $R_{0AB}$ are distances between two
dipolar excitons at the classical turning point for two A excitons,
two B excitons, and one A and one B excitons, correspondingly. Let
us mention that in the thermodynamical equilibrium the chemical
potentials of A and B dipolar excitons are equal.

From Eq.~(\ref{cond}) the following expressions are obtained
\begin{eqnarray}  \label{r0}
R_{0AA}=\left( \frac{ke^{2}D^{2}}{2\epsilon _{d}\left( \mu -\mathcal{A}%
_{A}\right) }\right) ^{1/3}\ ,\hspace{1cm}R_{0BB}=\left( \frac{ke^{2}D^{2}}{%
2\epsilon _{d}\left( \mu -\mathcal{A}_{B}\right) }\right) ^{1/3}\ ,\hspace{%
1cm}R_{0AB}=\left( \frac{ke^{2}D^{2}}{\epsilon _{d}\left( 2\mu -\mathcal{A}%
_{A}-\mathcal{A}_{B}\right) }\right) ^{1/3}\ .
\end{eqnarray}

Following the procedure presented in Ref.~\onlinecite{BKKL}, one can
obtain the interaction constants for the exciton-exciton interaction
\begin{eqnarray}  \label{g}
g_{AA}=\frac{2\pi ke^{2}D^{2}}{\epsilon _{d}R_{0AA}}\ ,\hspace{3cm}g_{BB}=%
\frac{2\pi ke^{2}D^{2}}{\epsilon _{d}R_{0BB}}\ ,\hspace{3cm}g_{AB}=\frac{%
2\pi ke^{2}D^{2}}{\epsilon _{d}R_{0AB}}\ .
\end{eqnarray}

 We expect that at zero temperature $T=0$ almost all A
and B excitons belong to the BEC of A and B excitons,
correspondingly. Therefore, we assume the formation of the binary
mixture of BECs. Using Bogoliubov approximation~\cite{Lifshitz},
generalized for two-component weakly-interacting Bose
gas~\cite{Timmermans}, we obtain the chemical potential $\mu $ of
the entire exciton system by minimizing $\hat{H}_{0}-\mu \hat{N}$
with respect to 2D concentration $n$, where $\hat{N}$ denotes the
number operator
\begin{eqnarray}  \label{Nop}
\hat{N}=\sum_{\mathbf{k}}a_{\mathbf{k}A}^{\dagger }a_{\mathbf{k}A}+\sum_{%
\mathbf{k}}a_{\mathbf{k}B}^{\dagger }a_{\mathbf{k}B}\ ,
\end{eqnarray}
and $H_{0}$ is the Hamiltonian describing the particles in the
condensate with zero momentum $\mathbf{k}=0$. In the Bologoiubov
approximation we
assume $N=N_{0}$, $a_{\mathbf{k}=0,A(B)}^{\dagger }=\sqrt{N_{0A(B)}}%
e^{-i\Theta _{A(B)}}$ and
$a_{\mathbf{k}=0,A(B)}=\sqrt{N_{0A(B)}}e^{i\Theta _{A(B)}}$, where
$N$ is the total number of all excitons, and $N_{0}$ is the number
of all excitons in the condensate, $N_{0A(B)}$ and $\Theta _{A(B)}$
are the number and phase for A(B) excitons in the corresponding
condensate. From Eqs.~(\ref{Ham}),~(\ref{Ham1}), and~(\ref{Ham2}) we
obtain
\begin{eqnarray}  \label{Ham3}
\hat{H}_{0}-\mu \hat{N}=S\left[ \left( \mathcal{A}_{A}-\mu \right)
n_{A}+\left( \mathcal{A}_{B}-\mu \right) n_{B}+\frac{g_{AA}n_{A}^{2}}{2}+%
\frac{g_{BB}n_{B}^{2}}{2}+g_{AB}n_{A}n_{B}\right] \ ,
\end{eqnarray}
where $n_{A}$ and $n_{B}$ are the 2D concentrations of A and B
excitons, correspondingly. The minimization of $\hat{H}_{0}-\mu
\hat{N}$ with respect to the number of A excitons $N_{A}=Sn_{A}$
results in
\begin{eqnarray}  \label{mu1}
\mu -\mathcal{A}_{A}=g_{AA}n_{A}+g_{AB}n_{B}\ .
\end{eqnarray}
The minimization of $\hat{H}_{0}-\mu \hat{N}$ with respect to the
number of B excitons $N_{B}=Sn_{B}$ results in
\begin{eqnarray}  \label{mu2}
\mu -\mathcal{A}_{B}=g_{BB}n_{B}+g_{AB}n_{A}\ .
\end{eqnarray}
From Eqs.~(\ref{mu1}) and~(\ref{mu2}), we obtain
\begin{eqnarray}  \label{mu3}
2\mu
-\mathcal{A}_{A}-\mathcal{A}_{B}=g_{AA}n_{A}+g_{BB}n_{B}+g_{AB}n\ ,
\end{eqnarray}
where $n=n_{A}+n_{B}$ is the total 2D concentration of excitons.

Combining
Eqs.~(\ref{r0}),~(\ref{g}),~(\ref{mu1}),~(\ref{mu2}),~(\ref{mu3}),
one obtains the following system of three cubical equations for the
interaction constants $g_{AA}$, $g_{BB}$, $g_{AB}$:
\begin{eqnarray}   \label{geq1}
&&g_{AA}^{3}-2\mathcal{B}n_{A}g_{AA}-2\mathcal{B}n_{B}g_{AB}=0\ ,
\nonumber
\\
&&g_{BB}^{3}-2\mathcal{B}n_{B}g_{BB}-2\mathcal{B}n_{A}g_{AB}=0\ ,
\\
&&g_{AB}^{3}-\mathcal{B}ng_{AB}-\mathcal{B}\left(
n_{A}g_{AA}+n_{B}g_{BB}\right) =0 \ ,  \nonumber
\end{eqnarray}%
where $B$ are defined as
\begin{eqnarray}  \label{AB}
\mathcal{B}=\frac{(2\pi )^{3}(ke^{2}D^{2})^{2}}{\epsilon_{d}^{2}}\ .
\end{eqnarray}%
Making the sum of the top two equations in~(\ref{geq1}), we can replace Eq.~(%
\ref{geq1}) by the following system of three cubical equations:
\begin{eqnarray}     \label{geq}
&&g_{AA}^{3}-2\mathcal{B}n_{A}g_{AA}-2\mathcal{B}n_{B}g_{AB}=0\ ,
\nonumber  \\
&&g_{BB}^{3}-2\mathcal{B}n_{B}g_{BB}-2\mathcal{B}n_{A}g_{AB}=0\ ,
\\
&&2g_{AB}^{3}=g_{AA}^{3}+g_{BB}^{3}\ . \nonumber
\end{eqnarray}%
The interaction constants $g_{AA}$, $g_{BB}$, $g_{AB}$ can be
obtained from
the solution of the system of three cubical equations represented by Eq.~(%
\ref{geq}).

If the interaction constants for exciton-exciton interaction are
negative, the spectrum of collective excitations at small momenta is
imaginary which reflects the instability of the excitonic ground
state~\cite{BKL_i,BKL_ii}. The system of equations Eq.~(\ref{geq})
has all real and positive roots only
if $g_{AA}=g_{BB}=g_{AB}\equiv g$. Substituting this condition into Eq.~(\ref{geq}%
), we obtain
\begin{eqnarray}  \label{geq222}
g^{3}-2\mathcal{B}\left( n_{A}+n_{B}\right) g=0\ .
\end{eqnarray}
Using $n=n_{A}+n_{B}$, we get from Eq.~(\ref{geq222}) the following
expression for $g$
\begin{eqnarray}  \label{geqeq}
g=\sqrt{2\mathcal{B}n}\ .
\end{eqnarray}
Substituting $\mathcal{B}$  from Eq.~(\ref{AB}) into
Eq.~(\ref{geqeq}), we obtain $g$ as
\begin{eqnarray}  \label{geqeq1}
g=\frac{4\pi ke^{2}D^{2}\sqrt{\pi n}}{\epsilon _{d}}\ .
\end{eqnarray}

Using the following notation,
\begin{eqnarray}  \label{not}
G_{AA} &=& g_{AA} n_{A} = g n_{A} , \hspace{2cm} G_{BB} = g_{BB} n_{B} = g
n_{B} , \hspace{2cm} G_{AB} = g_{AB}\sqrt{n_{A}n_{B}} = g \sqrt{n_{A}n_{B}}
\ ,  \nonumber \\
\omega_{A} (k) &=& \sqrt{\varepsilon_{(0)A}^{2}(k) + 2
G_{AA}\varepsilon_{(0)A}(k)} \ ,   \\
\omega_{B}(k) &=& \sqrt{\varepsilon_{(0)B}^{2}(k) + 2
G_{BB}\varepsilon_{(0)B}(k)} \ ,  \nonumber
\end{eqnarray}
we obtain two modes of the spectrum of Bose collective excitations $%
\varepsilon_{j}(k)$ in the Bogoliubov approximation for two-component
weakly-interacting Bose gas~\cite{Sun}
\begin{eqnarray}  \label{col}
\varepsilon_{j}(k) = \sqrt{\frac{\omega_{A}^{2}(k) + \omega_{B}^{2}(k) +
(-1)^{j-1}\sqrt{\left(\omega_{A}^{2}(k) - \omega_{B}^{2}(k)\right)^{2} +
\left(4G_{AB}\right)^{2}\varepsilon_{(0)A}(k)\varepsilon_{(0)B}(k)} }{2}} \ ,
\end{eqnarray}
where $j=1$, $2$. We can note that $G_{AB}^{2} = G_{AA}G_{BB}$.

In the limit of small momenta $p$, when $\varepsilon_{(0)A}(k) \ll
G_{AA}$ and $\varepsilon_{(0)B}(k) \ll G_{BB}$, we expand the spectrum of collective excitations $%
\varepsilon_{j}(k)$  up to the first order with respect to the
momentum $p = \hbar k$ and get two sound modes of the collective
excitations $\varepsilon_{j}(p) = c_{j}p$, where $c_{j}$ is the
sound velocity given by
\begin{eqnarray}  \label{c}
c_{j} = \sqrt{\frac{G_{AA}}{2M_{A}} + \frac{G_{BB}}{2M_{B}} + (-1)^{j-1}%
\sqrt{\left(\frac{G_{AA}}{2M_{A}} - \frac{G_{BB}}{2M_{B}}\right)^{2} + \frac{%
G_{AB}^{2}}{M_{A}M_{B}}}} \ ,
\end{eqnarray}

In the limit of large momenta, when $\varepsilon _{(0)A}(k)\gg G_{AA}$ and $%
\varepsilon _{(0)B}(k)\gg G_{BB}$, we get two parabolic modes of collective
excitations with the spectra $\varepsilon _{1}(k)=\varepsilon _{(0)A}(k)$
and $\varepsilon _{2}(k)=\varepsilon _{(0)B}(k)$, if $M_{A}<M_{B}$ and if $%
M_{A}>M_{B}$ with the spectra $\varepsilon _{1}(k)=\varepsilon
_{(0)B}(k)$ and $\varepsilon _{2}(k)=\varepsilon _{(0)A}(k)$ .

The Hamiltonian $\hat{H}_{col}$ of the collective excitations,
corresponding to two branches of the spectrum,  in the Bogoliubov
approximation for the entire two-component system is given
by~\cite{Sun}
\begin{eqnarray}  \label{Hamq}
\hat{H}_{col} = \sum_{\mathbf{k} \neq 0}\varepsilon_{1}(k)\alpha_{1\mathbf{k}%
}^{\dagger}\alpha_{1\mathbf{k}} + \sum_{\mathbf{k}\neq
0}\varepsilon_{2}(k)\alpha_{2\mathbf{k}}^{\dagger}\alpha_{2\mathbf{k}}
\ ,
\end{eqnarray}
where $\alpha_{j\mathbf{k}}^{\dagger}$ and $\alpha_{j\mathbf{k}}$ are the
creation and annihilation Bose operators for the quasiparticles with the
energy dispersion corresponding to the $j$th mode of the spectrum of the
collective excitations.

If A and B excitons do not interact, we put $g_{AB} = 0$ and $G_{AB} = 0$,
and in the limit of the small momenta we get for the sound velocity $c_{1} =
\sqrt{\frac{G_{AA}}{M_{A}}}$ and $c_{2} = \sqrt{\frac{G_{BB}}{M_{B}}}$,
which satisfies to the sound velocity in the Bogoliubov approximation for
one-component system~\cite{Lifshitz}.

If for simplicity we consider the specific case when the densities of A and
B excitons are the same $n_{A}=n_{B}=n/2$, we get from Eq.~(\ref{not})
\begin{eqnarray}
G_{AA} &=& G_{BB}=G_{AB}=gn/2\ ,  \nonumber  \label{not1} \\
\omega _{A}(k) &=&\sqrt{\varepsilon _{(0)A}^{2}(k)+gn\varepsilon _{(0)A}(k)}%
\ , \\
\omega _{B}(k) &=&\sqrt{\varepsilon _{(0)B}^{2}(k)+gn\varepsilon _{(0)B}(k)}%
\ .   \nonumber
\end{eqnarray}

From Eq.~(\ref{col}), we get the spectrum of collective excitations
\begin{eqnarray}  \label{col1}
\varepsilon _{j}(k)=\sqrt{\frac{\omega _{A}^{2}(k)+\omega
_{B}^{2}(k)+(-1)^{j-1}\sqrt{\left( \omega _{A}^{2}(k)-\omega
_{B}^{2}(k)\right) ^{2}+4g^{2}n^{2}\varepsilon _{(0)A}(k)\varepsilon
_{(0)B}(k)}}{2}}\ ,
\end{eqnarray}
and the sound velocity at $n_{A}=n_{B}=n/2$ is obtained as
\begin{eqnarray}  \label{c1}
c_{j}=\sqrt{\frac{gn}{2}\left( \frac{1}{2M_{A}}+\frac{1}{2M_{B}}+(-1)^{j-1}%
\sqrt{\left( \frac{1}{2M_{A}}-\frac{1}{2M_{B}}\right) ^{2}+\frac{1}{%
M_{A}M_{B}}}\right) }\ .
\end{eqnarray}

It follows from Eq.~(\ref{c1}), that there is only one non-zero sound
velocity at $n_{A}=n_{B}=n/2$ given by
\begin{eqnarray}  \label{c2}
c=\sqrt{\frac{gn}{2}\left( \frac{1}{M_{A}}+\frac{1}{M_{B}}\right) }\
.
\end{eqnarray}
 Interestingly enough, if for an one-component system
the sound velocity  is inversely proportional to the square root of
the mass of the exciton, $M_{A}^{-1/2}$, $M_{B}^{-1/2}$ one or the
other, for a two-component system it  is inversely proportional to
the square root of the reduced mass of two excitons, $\mu
_{AB}^{-1/2}$, where $\mu _{AB}=M_{A}M_{B}/(M_{A}+M_{B})$.  Since
$M_{A}>\mu _{AB}$ and $M_{B}>\mu
_{AB}$, it is  always true that  $M_{A}^{-1/2}<\mu _{AB}^{-1/2}$ or $%
M_{B}^{-1/2}<\mu _{AB}^{-1/2}.$ Thus, in the Bogoliubov
approximation the sound velocity in a two-component system is always
larger than in an one-component system.

\section{Superfluidity}

\label{sup}

Since at small momenta the energy spectrum of the quasiparticles in
 the  weakly-interacting gas of dipolar excitons is
soundlike, this system
satisfies to the Landau criterion for superfluidity~\cite{Lifshitz,Abrikosov}%
. The critical velocity for the superfluidity is given by $v_{c} =
\min\left(c_{1},c_{2}\right)$, because the quasiparticles are created at the
velocities above the velocity of sound for the lowest mode of the
quasiparticle dispersion.

The density of the superfluid component $\rho _{s}(T)$ is defined as
$\rho _{s}(T)=\rho -\rho _{n}(T)$, where $\rho
=M_{A}n_{A}+M_{B}n_{B}$ is the total 2D density of the system and
$\rho _{n}(T)$ is the density of the normal component. We define the
normal component density $\rho _{n}(T)$ by the standard procedure
\cite{Pitaevskii}. Suppose that the exciton system moves with a
velocity $\mathbf{u}$, which means that the superfluid component
moves with  the velocity $\mathbf{u}$. At nonzero temperatures $T$
dissipating quasiparticles will appear in this system. Since their
density is small at low temperatures, one can assume that the gas of
quasiparticles is an ideal Bose gas.  To calculate the superfluid
component density, we define the total mass current for a
two-component Bose-gas of quasiparticles in  the frame, in which the
superfluid component is at rest, as
\begin{eqnarray}  \label{nnor}
\mathbf{J}=\int \frac{d^{2}p}{(2\pi \hbar )^{2}}\mathbf{p}\left( f\left[
\varepsilon _{1}(p)-\mathbf{p}\mathbf{u}\right] +f\left[ \varepsilon _{2}(p)-%
\mathbf{p}\mathbf{u}\right] \right) \ ,
\end{eqnarray}
where $f\left[ \varepsilon _{1}(p))\right] =\left( \exp \left[
\varepsilon
_{1}(p)/(k_{B}T)\right] -1\right) ^{-1}$ and $f\left[ \varepsilon _{2}(p))%
\right] =\left( \exp \left[ \varepsilon _{2}(p)/(k_{B}T)\right]
-1\right) ^{-1}$ are the Bose-Einstein distribution function for the
quasiparticles with the dispersion $\varepsilon _{1}(p)$ and
$\varepsilon _{2}(p)$, correspondingly, $k_{B}$ is the Boltzmann
constant. Expanding the expression under the integral in terms of
$\mathbf{p}\mathbf{u}/(k_{B}T)$ and restricting ourselves by the
first order term, we obtain:
\begin{eqnarray}  \label{J_Tot}
\mathbf{J}=-\frac{\mathbf{u}}{2}\int \frac{d^{2}p}{(2\pi \hbar )^{2}}%
p^{2}\left( \frac{\partial f\left[ \varepsilon _{1}(p)\right] }{\partial
\varepsilon _{1}(p)}+\frac{\partial f\left[ \varepsilon _{2}(p)\right] }{%
\partial \varepsilon _{2}(p)}\right) \ .
\end{eqnarray}
The density $\rho _{n}$ of the normal component is defined as~\cite%
{Pitaevskii}
\begin{eqnarray}  \label{J_M}
\mathbf{J}=\rho _{n}\mathbf{u}\ .
\end{eqnarray}
Using Eqs.~(\ref{J_Tot}) and~(\ref{J_M}), we obtain the density of
the normal component as
\begin{eqnarray}  \label{rhon}
\rho _{n}(T)=-\frac{1}{2}\int \frac{d^{2}p}{(2\pi \hbar
)^{2}}p^{2}\left( \frac{\partial f\left[ \varepsilon _{1}(p)\right]
}{\partial \varepsilon _{1}(p)}+\frac{\partial f\left[ \varepsilon
_{2}(p)\right] }{\partial \varepsilon _{2}(p)}\right) \ .
\end{eqnarray}
At small temperatures $k_{B}T\ll M_{A(B)}c_{j}^{2}$, the small
momenta,
corresponding to the conditions $\varepsilon _{(0)A}(k)\ll G_{AA}$ and $%
\varepsilon _{(0)B}(k)\ll G_{BB}$ provide the main contribution to
the integral in the r.h.s. of Eq.~(\ref{rhon}), which corresponds to
the quasiparticles with the sound spectrum $\varepsilon
_{j}(k)=c_{j}k$ with the sound velocity given by Eq.~(\ref{c}),
results in
\begin{eqnarray}  \label{rhon1}
\rho _{n}(T)=\frac{3\zeta (3)}{2\pi \hbar ^{2}}k_{B}^{3}T^{3}\left( \frac{1}{%
c_{1}^{4}}+\frac{1}{c_{2}^{4}}\right) \ ,
\end{eqnarray}
where $\zeta (z)$ is the Riemann zeta function ($\zeta (3)\simeq
1.202$).

For high temperatures $k_{B}T\gg M_{A(B)}c_{j}^{2}$, the large momenta $%
M_{A(B)}c_{j}^{2}\ll \varepsilon _{0A(B)}(k)\ll k_{B}T$ provide the main
contribution to the integral in the r.h.s. of Eq.~(\ref{rhon}), which
corresponds to quasiparticles with the parabolic spectrum. Using the result
for these values of momenta for one-component system~\cite{Pitaevskii}, we
get for high temperatures
\begin{eqnarray}  \label{rhon2}
\rho _{n}(T)=\left\{
\begin{array}{c}
\frac{k_{B}T}{2\pi \hbar ^{2}}\left( M_{A}^{2}\ln \frac{k_{B}T}{%
M_{A}c_{1}^{2}}+M_{B}^{2}\ln \frac{k_{B}T}{M_{B}c_{2}^{2}}\right) \ ,\text{
if } M_{A}<M_{B} \\
\frac{k_{B}T}{2\pi \hbar ^{2}}\left( M_{A}^{2}\ln \frac{k_{B}T}{%
M_{A}c_{2}^{2}}+M_{B}^{2}\ln \frac{k_{B}T}{M_{B}c_{1}^{2}}\right) \
,\text{ if }M_{A}>M_{B} \ .
\end{array}%
\right.
\end{eqnarray}

Neglecting the interaction between the quasiparticles, the mean field
critical temperature $T_{c}$ of the phase transition related to the
occurrence of superfluidity is given by the condition $%
\rho_{s}(T_{c}) = 0$~\cite{Pitaevskii}:
\begin{eqnarray}  \label{Tc1}
\rho_{n} (T_{c}) = \rho = M_{A} n_{A} + M_{B}n_{B} \ .
\end{eqnarray}

At small temperatures $k_{B}T\ll M_{A(B)}c_{j}^{2}$ substituting Eq.~(\ref%
{rhon1}) into Eq.~(\ref{Tc1}), we get
\begin{eqnarray}  \label{Tc2}
T_{c}=\left[ \frac{2\pi \hbar ^{2}\rho }{3\zeta (3)k_{B}^{3}\left( \frac{1}{%
c_{1}^{4}}+\frac{1}{c_{2}^{4}}\right) }\right] ^{1/3}\ .
\end{eqnarray}
If $T_{c}$ obtained from Eq.~(\ref{Tc2}) satisfies to the condition $%
k_{B}T_{c}\ll M_{A(B)}c_{j}^{2}$, it is the right value of the mean
field critical temperature. Otherwise, at high temperatures
$k_{B}T\gg M_{A(B)}c_{j}^{2}$, we obtain the critical temperature
$T_{c}$ from the solution of the equation
\begin{eqnarray}  \label{Tc3}
\rho =\left\{
\begin{array}{c} \frac{k_{B}T_{c}}{2\pi \hbar ^{2}}\left( M_{A}^{2}\ln \frac{k_{B}T_{c}%
}{M_{A}c_{1}^{2}}+M_{B}^{2}\ln
\frac{k_{B}T_{c}}{M_{B}c_{2}^{2}}\right) \ ,\text{ if }M_{A}<M_{B} \\
\frac{k_{B}T_{c}}{2\pi \hbar ^{2}}\left( M_{A}^{2}\ln \frac{k_{B}T_{c}%
}{M_{A}c_{2}^{2}}+M_{B}^{2}\ln
\frac{k_{B}T_{c}}{M_{B}c_{1}^{2}}\right)   \ ,\text{ if
}M_{A}>M_{B}c \ .
\end{array}%
\right.
\end{eqnarray}

At $n_{A}=n_{B}=n/2$, we get the density of the normal component as
\begin{eqnarray}  \label{rhon1e}
\rho _{n}(T)=\left\{
\begin{array}{c} \frac{3\zeta (3)k_{B}^{3}T^{3}}{2\pi \hbar ^{2}c^{4}}  \ ,\text{
at low temperatures } \\
\frac{k_{B}T}{2\pi \hbar ^{2}}\left( M_{A}^{2}\ln \frac{k_{B}T}{%
M_{A}c^{2}}+M_{B}^{2}\ln \frac{k_{B}T}{M_{B}c^{2}}\right)  \ ,
\text{ at high temperatures. }
\end{array}%
\right.
\end{eqnarray}

At $n_{A}=n_{B}=n/2$, for the low-temperature case we get the mean
field critical temperature as
\begin{eqnarray}  \label{Tc22e}
T_{c}=\left[ \frac{2\pi \hbar ^{2}\rho c^{4}}{3\zeta
(3)k_{B}^{3}}\right] ^{1/3}    \ .
\end{eqnarray}

 At the first glance, Eq. ~(\ref{Tc2e}) is the same as for
one-component exciton gas. However, after consideration of (45), one
obtains
\begin{eqnarray}  \label{Tc2e}
T_{c}=\left[ \frac{\pi \hbar ^{2}g^{2}n^{3}}{12\zeta (3)}Q\right]
^{1/3} \ .
\end{eqnarray}
where the parameter $Q$ is defined as
\begin{eqnarray}  \label{Qdef}
Q=\frac{M_{A}+M_{B}}{\left( \mu _{AB}\right)^{2}} \ ,
\end{eqnarray}
and $\mu _{AB}$ is the reduced mass for two-component system of A
and B excitons. For one-component dilute exciton gas $Q_{A}=1/M_{A}$
or $Q_{B}=1/M_{B},$ that is always less than the value of $Q$ for
 a two-component Bose gas of  A and B dipolar excitons.
Therefore, $T_{c}$ is always  higher for a two-component dilute
dipolar exciton gas than for an one-component dilute dipolar exciton
gas.

If $T_{c}$ obtained from Eq.~(\ref{Tc2e}) satisfies to the condition $%
k_{B}T_{c} \ll M_{A(B)}c^{2}$, it is the right value of the mean field
critical temperature. Otherwise, at high temperatures $k_{B}T \gg
M_{A(B)}c^{2}$, we obtain the critical temperature $T_{c}$ from the solution
of the equation
\begin{eqnarray}  \label{Tc3e}
\rho = \frac{k_{B}T_{c}}{2\pi\hbar^{2}}\left(M_{A}^{2}\ln\frac{k_{B}T_{c}}{%
M_{A}c^{2}} + M_{B}^{2}\ln\frac{k_{B}T_{c}}{M_{B}c^{2}} \right) \ ,
\end{eqnarray}
where $\rho = \left(M_{A} + M_{B}\right)n/2$.

\section{Two-component direct exciton superfluidity in a TMDC monolayer}

\label{directex}

Let us consider the two-component  weakly interacting Bose gas of
direct A and B excitons in a single TMDC monolayer. The direct
excitons of type A are formed by spin-up electrons from conduction
and spin-down holes from valence bands in a single TMDC monolayer.

There are two differences between two-component  weakly interacting
Bose gas of A and B excitons in a single TMDC monolayer and two
parallel TMDC layers with the spatially separated charge carriers.
The first difference is that the effective mass of direct excitons
in a single TMDC monolayer is different from the effective mass of
indirect excitons in two parallel TMDC layers given by
Eq.~(\ref{mex}). The second difference is that the interaction
constant for the contact exciton-exciton repulsion for direct
excitons in a single TMDC monolayer is different from the
interaction constant for the dipole-dipole exciton-exciton repulsion
for dipolar excitons in two parallel TMDC layers given by
Eq.~(\ref{geqeq1}).

As discussed in Refs.~\onlinecite{Ciuti-exex}
and~\onlinecite{Laikht} for a dilute exciton gas, the excitons can
be treated as bosons with a repulsive contact interaction. For small
wave vectors $q\ll \rho ^{-1}$ the exciton-exciton interaction
constant, describing the pairwise exciton-exciton repulsion between
A and A (B and B) direct excitons, correspondingly, can be
approximated by a contact potential
\begin{eqnarray}  \label{dirgaa}
g_{AA(BB)}=\frac{6ke^{2}a_{A(B)}}{\varepsilon _{m}}\ ,
\end{eqnarray}
where $a_{A(B)}$ is the exciton Bohr radius for A(B) direct
excitons, correspondingly. This direct exciton Bohr radius
$a_{A(B)}$ can be obtained analagousely to $\tilde{\beta}$ in
Eq.~(39) in Ref.~\onlinecite{BKZ} for a gapped graphene monolayer.
In Eq.~(\ref{dirgaa}), $\varepsilon _{m}$ is the dielectric constant
for the media, surrounding the TMDC monolayer, and for a freely
suspended TMDC material in vacuum we have $\varepsilon _{m}=1$. This
approximation for the exciton-exciton repulsion is applicable,
because resonantly excited excitons have very small wave
vectors~\cite{Ciuti}.

For the interaction constant, describing the pair contact repulsion
between A and B excitons, we use
\begin{eqnarray}  \label{dirgaa1}
g_{AB}=\frac{6ke^{2}a_{AB}}{\varepsilon _{m}}\ ,
\end{eqnarray}
where $a_{AB}$ is the phenomenological parameter. Assuming the value of $%
a_{AB}$ is the average of $a_{A}$ and $a_{B}$, we have
\begin{eqnarray}  \label{aabpar}
a_{AB}=\frac{a_{A}+a_{B}}{2}\ .
\end{eqnarray}

For direct excitons in a single TMDC monolayer, substituting the direct
exciton effective masses $M_{A(B)}$ and the direct excitons interaction
parameters $g_{AA(BB)}$ and $g_{AB}$ into Eqs.~(\ref{col}) and~(\ref{c}), we
obtain the two branches of the spectrum of collective excitations and the
sound velocities for direct excitons in a single TMDC monolayer. Then
substituting the sound velocities into Eqs.~(\ref{rhon1}) and~(\ref{Tc2}),
we obtain the density of the superfluid component as a function of
temperature and the mean-field temperature of the superfluid phase
transition, correspondingly for two-component weakly-interacting Bose gas of
direct excitons in as single TMDC monolayer.

The approach presented in this section can be easily applied to study the
two-component superfluidity of A and B exciton polaritons in a TMDC layer
embedded in a microcavity, which was studied in the experiment~\cite{Menon}.

\section{Results and Discussion}

\label{disc}

In this section, we present the results of our calculations. Since
the dipolar excitons were experimentally observed in two TMDC layers
separated by
 hBN insulating layers~\cite{Calman}, we assume in our calculations the
dielectric constant of the insulating barrier is the same as for hBN: $%
\varepsilon _{d}=7.1$. In our calculations we use the parameters $a,$ $t$, $%
\Delta $, and $\lambda $ for transition metal dichalcogenides $\mathrm{%
MoS_{2}}$, $\mathrm{MoSe_{2}}$, $\mathrm{WS_{2}}$, and $\mathrm{WSe_{2}}$
that are listed in Table 1 in Ref.~\onlinecite{Yao} and for $\mathrm{MoTe_{2}%
}$ and $\mathrm{WTe_{2}}$ from Ref.~\onlinecite{Kormanyos}. The
results of calculations for the effective masses of A and B excitons
for the layer separation $D=5\ \mathrm{nm}$ obtained from
Eq.~(\ref{mex}) are represented in Table 1.

\begin{table}[tbp]
\caption{Effective masses of A and B excitons for different TMDC
materials in units of the free electron mass at the interlayer
separation $D = 5 \ \mathrm{nm}$.} \label{tab1}\centering
\begin{tabular}{ccccccc}
\hline\hline
& \multicolumn{6}{c}{Mass of exciton} \\ \hline
Exciton type & MoS$_{2}$ & MoSe$_{2}$ & MoTe$_{2}$ & WS$_{2}$ & WSe$_{2}$ &
WTe$_{2}$ \\ \hline
A & 0.499 & 0.555 & 0.790 & 0.319 & 0.345 & 0.277 \\
B & 0.545 & 0.625 & 0.976 & 0.403 & 0.457 & 0.501 \\ \hline\hline
\end{tabular}%
\end{table}

According to  Table 1, the B excitons are heavier than the A
excitons for all TMDC. There is an advantage of our analytical
approach that illustrates the dependence of the effective exciton
masses on spin-orbit coupling resulting in the formation of two
types of excitons A and B in TMDC and their dependence on the
parameters $a,$ $t$, and $\Delta $. Also the results of our
calculations show
 that the exciton effective mass very slightly depends on
the distance between two parallel TMDC layers $D.$




\begin{figure}[h]
\includegraphics[width=12.0cm]{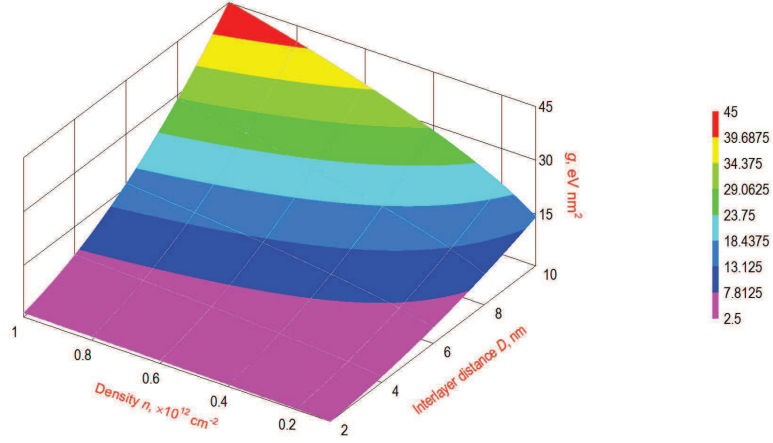}
\caption{The interaction constant $g$.} \label{gdfig}
\end{figure}

The interaction constant $g$ as a function of the interlayer
separation $D$ and exciton concentration $n$ is represented in
Fig.~\ref{gdfig}. According to Fig.~\ref{gdfig}, the effective
interaction constant $g $ increases with the increase of the
interlayer separation $D$ and the increase of the exciton
concentration $n$.

The mean field critical temperature $T_{c}$ for the excitonic
superfluidity obtained from Eq.~(\ref{Tc2e}) is represented in
Table~2. The critical temperature $T_{c}$ was calculated for
different TMDC at the moderated concentration of excitons $n=3\times
10^{11}$ cm$^{-2}$, assuming $n_{A}=n_{B} = n/2$.

\begin{table}[tbp]
\caption{Effective and reduced masses and factor $Q$ for
two-component exciton gas of different TMDC materials at the
interlayer separation $D = 5 \ \mathrm{nm}$.} \label{tab2}\centering
\begin{tabular}{ccccccc}
\hline\hline
& MoS$_{2}$ & MoSe$_{2}$ & MoTe$_{2}$ & WS$_{2}$ & WSe$_{2}$ & WTe$_{2}$ \\
\hline
$M_{A}+M_{B}$ & 1.044 & 1.180 & 1.766 & 0.722 & 0.802 & 0.778 \\
$\mu _{AB}$ & 0.261 & 0.294 & 0.437 & 0.178 & 0.197 & 0.179 \\
$Q$ & 15.380 & 13.655 & 9.260 & 22.750 & 20.769 & 24.453 \\ \hline\hline
\end{tabular}%
\end{table}

\begin{table}[tbp]
\caption{Dependence of the mean field critical temperature $T_{c}$
on the interlayer separation $D$ for different TMDC materials.}
\label{tab3}\centering
\begin{tabular}{ccccccc}
\hline\hline
$D$, nm & \multicolumn{6}{c}{$T_{c},$ K} \\ \hline
& MoS$_{2}$ & MoSe$_{2}$ & MoTe$_{2}$ & WS$_{2}$ & WSe$_{2}$ & WTe$_{2}$ \\
2 & 22 & 21 & 19 & 25 & 24 & 26 \\
3 & 38 & 36 & 32 & 43 & 41 & 44 \\
4 & 55 & 53 & 47 & 63 & 61 & 64 \\
5 & 74 & 71 & 63 & 85 & 82 & 87 \\
6 & 95 & 91 & 80 & 108 & 105 & 110 \\
7 & 116 & 112 & 98 & 132 & 128 & 136 \\ \hline\hline
\end{tabular}%
\end{table}

Let us mention that for our calculations we substituted the exciton
concentration $n=3\times 10^{11}$ cm$^{-2}$ smaller than the maximal
exciton concentration obtained in the experiment~\cite{You_bi}:
$n_{max}=5\times 10^{11}$ cm$^{-2}$. The exciton concentration
$n=3\times 10^{11}$ cm$^{-2}$ used in our calculations corresponds
to the degenerate exciton Bose gas in the phase
diagram~\cite{Fogler}. While, in general, the electron-hole
interaction is described by Keldysh's potential~\cite{Keldysh}, we
performed our calculations at the interlayer separation from $D = 2
\ \mathrm{nm}$ up to $D = 10 \ \mathrm{nm}$, when the screening
effects are negligible, and the electron-hole interaction is
described by Coulomb's potential. We used for our calculations the
interlayer separations $D$ larger than experimental
values~\cite{Calman} by the following reasons: (i) the larger $D$
leads to the increase of the potential barrier for electron-hole
tunneling between the layers, which results in the increase of the
exciton life-time; (ii) the larger $D$ leads to the increase of the
exciton dipole moment, which causes the increase of the
exciton-exciton dipole-dipole repulsion, and, therefore, the
increase of the sound velocity and the superfluid density, which
results in the increase of the mean field temperature of the
superfluidity, which can be seen in the Table~2.

\begin{figure}[h]
\includegraphics[width=12.0cm]{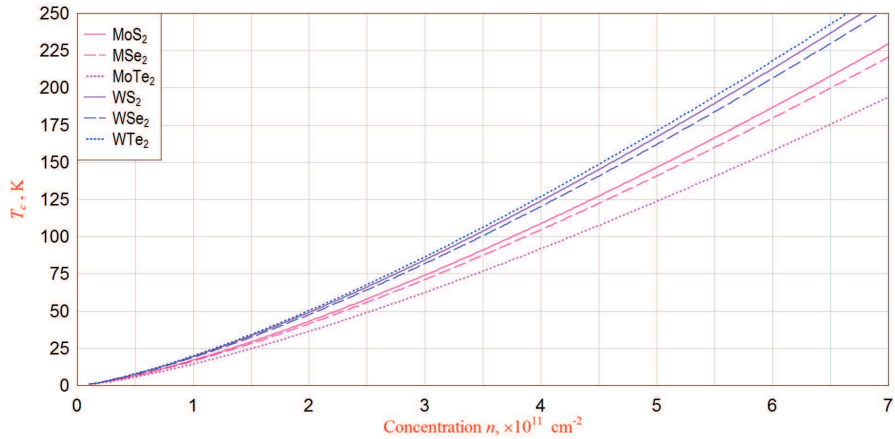}
\caption{The mean field critical temperature of the superfluidity
$T_{c}$ as a function of the exciton concentration $n$ for different
TMDC materials  at the interlayer separation $D = 5 \ \mathrm{nm}$.}
\label{tcfigAll}
\end{figure}

The mean field critical temperature of the superfluidity $T_{c}$ as
a function of the exciton concentration $n$ for MX$_{2}$ materials
is represented in Fig.~\ref{tcfigAll}. According to
Fig.~\ref{tcfigAll}, the critical temperature $T_{c}$ increases with
the increase of the exciton concentration $n$, and $T_{c}$ is
increased for different TMDC materials in
the following order: $\mathrm{MoTe_{2}}$, $\mathrm{MoSe_{2}}$, $\mathrm{%
MoS_{2}}$, $\mathrm{WSe_{2}}$, $\mathrm{WS_{2}}$,
$\mathrm{WTe_{2}}$. The critical temperature $T_{c}$ for all
chalcogenides $\mathrm{Se}$, $\mathrm{S}
$ and $\mathrm{Te}$ is larger for $\mathrm{WX_{2}}$ than for $\mathrm{MoX_{2}%
}$. It can be noticed that the order of types of different TMDC
materials with respect to the increase of $T_{c}$, presented in
Table 3 and Fig.~\ref{tcfigAll}, is exactly the same as the order of
these TMDC materials with respect to the increase  of the parameter
$Q$, presented in Table 2. This is caused by the fact that,
according to Eq.~(\ref{Tc2e}), $T_{c}$ is directly proportional to
$Q^{1/3}$. Let us mention that the order of types of different TMDC
materials with respect to the increase of $T_{c}$ is different from
the order of these materials with respect to the exciton effective
masses, presented in Table 1, and the parameters $a$, $t$, and the
separation between $X$ planes $d_{X-X}$ taken from
Ref.~\onlinecite{Kormanyos}.

The exciton-exciton interaction for studied in this Paper
two-component weakly-interacting dilute system of A and B dipolar
excitons leads to two branches of collective excitation spectrum
characterized at small momenta by two different sound velocities
$c_{1}$ and $c_{2}$, which is caused by the fact that the system
under consideration is a two-component  system. Therefore, at small
temperatures the normal component is formed by the contributions
from two types of quasiparticles, corresponding to two different
branches of the collective excitations spectra with two different
sound velocities at  small momenta. All general expressions for the
density of the normal and superfluid components and the mean-field
phase transition temperature were calculated in this Paper, taking
into account the existence of two branches of the spectrum of
collective excitations. The calculations were presented for the
specific case, when the concentrations of A and B excitons are
equal, and the collective spectrum is characterized at small momenta
by only one non-zero sound velocity.

We conclude that the critical temperature $T_{c}$ for superfluidity
for two-component exciton  gas in  a TMDC bilayer is about one order
of magnitude higher than $T_{c}$ for one-component exciton  gas
 in the semiconductor coupled quantum wells. According to
Eq.~(\ref{Tc2e}), the mean field critical temperature $T_{c}$ is
directly proportional to the parameter $Q^{1/3}$, which is
determined by the exciton reduced mass $\mu _{AB}$ and the sum of A
and B exciton masses $M_{A}+M_{B}$, while for the one-component
exciton Bose gas in CQWs $Q=M^{-1/3}$, where $M$ is the exciton mass
in CQWs. For example,  if $M_{A}$=$M_{B}$=$M$, then  $Q$ for an
one-component gas is eight times less than for a two-component Bose
gas. Thus, for one-component dilute exciton gas $Q$ is always less
than the value of $Q$ for a two-component Bose gas of A and B
dipolar excitons. It can be easily seen that the inequalities $\mu
_{AB}<M_{A}+M_{B}$ and $Q^{1/3}>\left( M_{A}+M_{B}\right) ^{-1/3}$
are always true for any positive $M_{A}$ and $M_{B}$. Therefore,
$T_{c}$ is always higher for a two-component dilute dipolar exciton
gas than for any one-component dilute dipolar exciton gas in
semiconductor CQWs in spite of the fact that the exciton masses for
A and B excitons in CQWs are the same order of magnitude as exciton
masses in CQWs. The advantage of the superfluidity of the dipolar
excitons in
 a TMDC bilayer is in the possibility of the creating
the superconducting electric currents in each TMDC layer by applying
the external voltage, since electrons and holes in each monolayers
are charge carriers. Since the quasiparticle gap and the exciton
binding energy in TMDC can be tuned by externally applied voltage
\cite{Chernikov_ef}, the effective mass, the sound velocity for the
collective excitations, the density of the superfluid component and
the phase transition temperature for superfluidity can also be
controlled by externally applied voltage.

\section{Conclusions}

\label{conc}

We propose a physical realization to observe high-temperature
superconducting electron-hole currents in two parallel TMDC layers
which is caused by the superfluidity of quasi-two-dimensional
dipolar A and B excitons in  a TMDC bilayer. The effective exciton
mass for A and B excitons is calculated analytically. The spectrum
of collective excitations obtained in the Bogoliubov approximation
for TMDC bilayer is characterized by two branches, reflecting the
fact that the exciton system under consideration is a two-component
weakly interacting Bose gas of A and B excitons. Two sound
velocities for both branches of the collective spectrum are derived
for two-component dipolar exciton system. It is shown that in the
Bogolubov approximation the sound velocity in a two-component system
is always larger than in an one-component system. The superfluid
density, defined by the contributions from the collective
excitations from two branches of collective spectrum, is obtained as
a function of temperature for two-component system of A and B
dipolar excitons. We show that the superfluid density and the
mean-field phase transition temperature for superfluid increase with
the increase of the excitonic concentration. The mean field critical
temperature for the phase transition is analyzed for various TMDC
materials. The mean field phase transition temperature, calculated
for dipolar exciton bilayer, is about one order of magnitude higher
than for any one-component exciton system of semiconductor CQWs due
to the fact that $T_{c}$ for two-component exciton system in TMDC
depends on the exciton reduced mass for the two-component system of
A and B excitons, more exactly, depends on the factor $Q$, which is
much larger for a two-component system than for an one-component
exciton system.

\acknowledgments

 The authors are grateful to A. Chernikov, M. Hybertsen, A. Moran, D.
Snoke for the valuable and stimulating discussions. This work was supported by NSF grant 1547751.

\appendix

\section{Eigenvalue problem for two particles}

\label{app:A}

Let us introduce the following notations:
\begin{eqnarray}
\mathcal{K}_{+} &=&\mathcal{K}\mathtt{e}^{i\Theta }=\mathcal{K}_{x}+i%
\mathcal{K}_{y}\ ,  \nonumber \\
\mathcal{K}_{-} &=&\mathcal{K}\mathtt{e}^{-i\Theta }=\mathcal{K}_{x}-i%
\mathcal{K}_{y}\ ,  \nonumber \\
\Theta  &=&\tan ^{-1}\left( {\frac{\mathcal{K}_{y}}{\mathcal{K}_{x}}}\right)
\ ,
\end{eqnarray}
and represent the Hamiltonian~(\ref{k20}) in the form of a $2\times 2$
matrix as
\begin{eqnarray}
\mathcal{H}_{\uparrow (\downarrow )}=\left( {\
\begin{array}{cc}
\mathcal{O}_{2}+V(r)\sigma _{0}-\frac{\Delta ^{\prime }}{2}\sigma _{0}+\frac{%
\Delta ^{\prime }}{2}\sigma _{3} & \mathcal{O}_{1} \\
\mathcal{O}_{1}^{\dagger } & \mathcal{O}_{2}+V(r)\sigma _{0}+\frac{\Delta
^{\prime }}{2}\sigma _{0}+\frac{\Delta ^{\prime }}{2}\sigma _{3}%
\end{array}%
}\right) \ ,  \label{22ham}
\end{eqnarray}
where $\mathcal{O}_{1}$ and $\mathcal{O}_{2}$ are given by
\begin{eqnarray}
\mathcal{O}_{1} &=&at\left( {\ \alpha \mathcal{K}_{-}-i\partial
_{x}-\partial _{y}}\right) \sigma _{0},  \label{oo} \\
\mathcal{O}_{2} &=&-at\left( {\
\begin{array}{cc}
0 & \beta \mathcal{K}_{-}+i\partial _{x}+\partial _{y} \\
\beta \mathcal{K}_{+}+i\partial _{x}-\partial _{y} & 0%
\end{array}%
}\right) .  \label{000}
\end{eqnarray}%
In Eqs. (A3) and (A4) $x$ and $y$ are the components of vector $\mathbf{r}$,
$\sigma _{j}$ are the Pauli matrices, $\sigma _{0}$ is the $2\times 2$ unit
matrix.

The eigenvalue problem~(\ref{l}) for the Hamiltonian~(\ref{22ham}) results
in the following coupled equations:
\begin{eqnarray}
\left( {\mathcal{O}_{2}+V(r)\sigma _{0}-\frac{\Delta ^{\prime }}{2}\sigma
_{0}+\frac{\Delta ^{\prime }}{2}\sigma _{3}}\right) \Psi _{c\uparrow
(\downarrow )}+\mathcal{O}_{1}\Psi _{v\uparrow (\downarrow )} &=&\epsilon
_{\uparrow (\downarrow )}\sigma _{0}\Psi _{c\uparrow (\downarrow )}
\nonumber \\
\mathcal{O}_{1}^{\dagger }\Psi _{c\uparrow (\downarrow )}+\left( {\mathcal{O}%
_{2}+V(r)\sigma _{0}+\frac{\Delta ^{\prime }}{2}\sigma _{0}+\frac{\Delta
^{\prime }}{2}\sigma _{3}}\right) \Psi _{v\uparrow (\downarrow )}
&=&\epsilon _{\uparrow (\downarrow )}\sigma _{0}\Psi _{v\uparrow (\downarrow
)}\ .  \label{221}
\end{eqnarray}
It follows from Eq.~(\ref{221}) that
\begin{eqnarray}
\Psi _{v\uparrow (\downarrow )}=\left( {\epsilon _{\uparrow
(\downarrow
)}\sigma _{0}-\mathcal{O}_{2}-V(r)\sigma _{0}-\frac{\Delta ^{\prime }}{2}%
\sigma _{0}-\frac{\Delta ^{\prime }}{2}\sigma _{3}}\right) ^{-1}\mathcal{O}%
_{1}^{\dagger }\Psi _{c\uparrow (\downarrow )}\ .    \label{a2}
\end{eqnarray}
Assuming the electron-hole attraction potential energy and both
relative and center-of-mass kinetic energies are small compared to
the gap $\Delta^{\prime }$, the following approximation is applied:
\begin{eqnarray}  \label{apsimi}
\left( \epsilon _{\uparrow (\downarrow )}\sigma _{0}-\mathcal{O}%
_{2}-V(r)\sigma _{0}-\frac{\Delta ^{\prime }}{2}\sigma _{0}-\frac{\Delta
^{\prime }}{2}\sigma _{3}\right) ^{-1}\simeq (\epsilon _{\uparrow
(\downarrow )}\sigma _{0}-\frac{\Delta ^{\prime }}{2}\sigma _{0}-\frac{%
\Delta ^{\prime }}{2}\sigma _{3})^{-1} \ .
\end{eqnarray}

Applying
\begin{equation}
\mathcal{O}_{1}^{\dagger}\mathcal{O}_{1}=a^{2}t^{2}\left( {\alpha ^{2}%
\mathcal{K}^{2}-\nabla _{\mathbf{r}}^{2}-2i\alpha (\mathcal{K}_{x}\partial
_{y}+\mathcal{K}_{y}\partial _{x})}\right) \sigma _{0}\ ,
\end{equation}%
and using Eq.~(\ref{wave function1}) we obtain from Eq.~(\ref{221}) for the
individual spinor components the following equations:
\begin{eqnarray}
&& \left[V(r)+ \frac{a^{2}t^{2}\left(\alpha^{2}\mathcal{K}^{2}-\nabla _{%
\mathbf{r}}^{2}-2i\alpha (\mathcal{K}_{x}\partial _{x}+\mathcal{K}%
_{y}\partial _{y})\right)}{\epsilon_{\uparrow(\downarrow)} - \Delta^{\prime}}%
\right] \phi_{c\uparrow(\downarrow)c\uparrow(\downarrow)}  \nonumber \\
&-& at \left( {\beta \mathcal{K}_{-}+i\partial _{x}+\partial _{y}}\right)
\phi_{c\uparrow(\downarrow)v\uparrow(\downarrow)}
=\epsilon_{\uparrow(\downarrow)}
\phi_{c\uparrow(\downarrow)c\uparrow(\downarrow)} \ ,  \label{231}
\end{eqnarray}%
\begin{eqnarray}
&& - at \left( {\ \beta \mathcal{K}_{+}+i\partial _{x}-\partial _{y}}\right)
\phi_{c\uparrow(\downarrow)c\uparrow(\downarrow)}  \nonumber \\
&+&\left[ V(r) - \Delta^{\prime} + \frac{a^{2}t^{2}\left(\alpha ^{2}\mathcal{%
K}^{2}-\nabla _{\mathbf{r}}^{2}-2i\alpha \left(\mathcal{K}_{x}\partial _{x}+%
\mathcal{K}_{y}\partial _{y}\right)\right)}{\epsilon_{\uparrow(\downarrow)}} %
\right] \phi_{c\uparrow(\downarrow)v\uparrow(\downarrow)}=\epsilon_{%
\uparrow(\downarrow)} \phi_{c\uparrow(\downarrow)v\uparrow(\downarrow)} \ .
\label{a7}
\end{eqnarray}%

Following the procedure applied for calculation of the energy spectrum of
the indirect excitons formed in two parallel gapped graphene layers~\cite%
{BKZ}, one gets from Eq.~(\ref{231}) for the spinor component
\begin{eqnarray}  \label{21}
\phi_{c\uparrow(\downarrow)c\uparrow(\downarrow)} = -\left(
\epsilon_{\uparrow(\downarrow)} - V(r) - \frac{a^{2}t^{2}\left(\alpha^{2}%
\mathcal{K}^{2}-\nabla_{\mathbf{r}}^{2} - 2i\alpha\left(\mathcal{K}%
_{x}\partial_{x} + \mathcal{K}_{y}\partial_{y}\right)\right)}{%
\epsilon_{\uparrow(\downarrow)} - \Delta^{\prime}}\right)^{-1}\left(at\left(%
\beta\mathcal{K}_{-} + i\partial_{x}+
\partial_{y}\right)\phi_{c\uparrow(\downarrow)v\uparrow(\downarrow)}\right)\
.
\end{eqnarray}

Assuming that the interaction potential and both the relative and
center-of-mass kinetic energies are small compared to the exciton energy, we
apply the following approximation:
\begin{eqnarray}  \label{ap}
\left[\epsilon_{\uparrow(\downarrow)} - V(r) - \frac{a^{2}t^{2}\left(%
\alpha^{2}\mathcal{K}^{2}-\nabla_{\mathbf{r}}^{2} - 2i\alpha\left(\mathcal{K}%
_{x}\partial_{x} + \mathcal{K}_{y}\partial_{y}\right)\right)}{%
\epsilon_{\uparrow(\downarrow)} - \Delta^{\prime}}\right]^{-1} \approx \frac{%
1}{\epsilon_{\uparrow(\downarrow)}} \ .
\end{eqnarray}

Substituting $\phi_{c\uparrow(\downarrow)c\uparrow(\downarrow)}$ from Eq.~(%
\ref{21}) into Eq.~(\ref{a7}) and applying the approximation given by Eq.~(%
\ref{ap}), we obtain
\begin{eqnarray}  \label{24}
&& \left[ -\Delta^{\prime} + V(r) + \frac{a^{2}t^{2}\left(\beta^{2}\mathcal{K%
}^{2} - \nabla_{\mathbf{r}}^{2} + 2i \beta \left(\mathcal{K}_{x}\partial_{x}
+ \mathcal{K}_{y} \partial_{y}\right)\right)}{\epsilon_{\uparrow(\downarrow)}%
} + \frac{a^{2}t^{2}\left(\alpha^{2}\mathcal{K}^{2}-\nabla_{\mathbf{r}}^{2}
- 2i\alpha\left(\mathcal{K}_{x}\partial_{x} + \mathcal{K}_{y}\partial_{y}%
\right)\right)}{\epsilon_{\uparrow(\downarrow)}}\right]\phi_{c\uparrow(%
\downarrow)v\uparrow(\downarrow)}  \nonumber \\
&=& \epsilon_{\uparrow(\downarrow)}
\phi_{c\uparrow(\downarrow)v\uparrow(\downarrow)} \ .
\end{eqnarray}

Choosing the values for the coefficients $\alpha$ and $\beta$ to separate
the coordinates of the center-of-mass (the wave vector $\mathcal{K}$ ) and
relative motion $\mathbf{r}$ in Eq.~(\ref{24}), we have
\begin{eqnarray}  \label{albe}
\alpha = \frac{1}{2} \ , \hspace{5cm} \beta = \frac{1}{2} \ .
\end{eqnarray}

Substituting Eq.~(\ref{albe}) into Eq.~(\ref{24}), we get
\begin{eqnarray}  \label{25}
\left[ - \frac{2a^{2}t^{2}\nabla_{\mathbf{r}}^{2}}{\epsilon_{\uparrow(%
\downarrow)}} + \frac{a^{2}t^{2}\mathcal{K}^{2}}{2\epsilon} -
\Delta^{\prime} + V(r)\right] \phi_{c\uparrow(\downarrow)v\uparrow(%
\downarrow)} = \epsilon_{\uparrow(\downarrow)}
\phi_{c\uparrow(\downarrow)v\uparrow(\downarrow)} \ .
\end{eqnarray}

\section{Solution of the equation for the single exciton spectrum}

\label{app:B}

Introducing $x = \sqrt{\epsilon_{\uparrow(\downarrow)}}$, we present Eq.~(%
\ref{geneq}) in the following form
\begin{eqnarray}  \label{geneqx}
2x^{4} + 2\left(\Delta^{\prime} + V_{0}\right) x^{2} - \frac{8atN\sqrt{\gamma%
}x}{\sqrt{2}} - a^{2} t^{2} \mathcal{K}^{2} = 0 \ .
\end{eqnarray}

For small momenta $\hbar \mathcal{K}$, we assume
\begin{eqnarray}  \label{xdx}
x = x_{0} + \Delta x \ ,
\end{eqnarray}
where $x=x_{0}$ corresponds to $\mathcal{K} = 0$. In this case, we obtain
from Eq.~(\ref{geneqx}) the following equation:
\begin{eqnarray}  \label{geneqx0}
x_{0}^{3} + \left(\Delta^{\prime} + V_{0}\right) x_{0} - \frac{4atN\sqrt{%
\gamma}}{\sqrt{2}} = 0 \ .
\end{eqnarray}
The cubic equation (\ref{geneqx0}) has the following real root:
\begin{eqnarray}  \label{x0}
x_{0} = \left(\frac{q}{2} + \sqrt{\frac{q^{2}}{4} + \frac{p^{3}}{27}}%
\right)^{1/3} + \left(\frac{q}{2} - \sqrt{\frac{q^{2}}{4} + \frac{p^{3}}{27}}%
\right)^{1/3} \ ,
\end{eqnarray}
where the parameters $p$ and $q$ are given by
\begin{eqnarray}  \label{pq}
p = \Delta^{\prime} + V_{0} \ , \hspace{5cm} q = \frac{4atN\sqrt{\gamma}}{%
\sqrt{2}} \ .
\end{eqnarray}

Substituting Eq.~(\ref{xdx}) into Eq.~(\ref{geneqx}), in the first order
with respect to $\Delta x$, we obtain
\begin{eqnarray}  \label{dxc}
\Delta x=\frac{a^{2}t^{2}\mathcal{K}^{2}}{2C_{A(B)}}\ ,\hspace{3cm}%
C_{A(B)}=4x_{0}^{3}+2\left( \Delta ^{\prime }+V_{0}\right) x_{0}-\frac{4atN%
\sqrt{\gamma }}{\sqrt{2}}\ ,
\end{eqnarray}
where $C_{A}$ and $C_{B}$  are  related to the A and B excitons,
when $\Delta ^{\prime }=\Delta -\lambda $ and $\Delta ^{\prime
}=\Delta +\lambda $  are used for spin-down and spin-up particles,
respectively. Let us also mention that the value of $x_{0}$ and
$\Delta x$ are different for different TMDC material due to the
values of parameters $a,$ $t$, $\Delta ,$ and $\lambda .$ Using
Eq.~(\ref{geneqx0}), we simplify Eq.~(\ref{dxc}) as
\begin{eqnarray}  \label{dxc2}
C_{A(B)}=3x_{0}^{3}+\left( \Delta ^{\prime }+V_{0}\right) x_{0}\ .
\end{eqnarray}

Then we get for the exciton energy $\epsilon$ for A(B) exciton in
the first order with respect to $\Delta x$:
\begin{eqnarray}  \label{exnm}
\epsilon_{A(B)} = \epsilon_{\uparrow(\downarrow)} = \left(x_{0} +
\Delta x\right)^{2} \approx x_{0}^{2} + 2x_{0}\Delta x \ ,
\end{eqnarray}
where $x_{0}$ has the different value  for A and B excitons.

\end{document}